\documentclass[11pt]{article}%
\usepackage{amsfonts}
\usepackage{amssymb}
\usepackage{graphicx}
\usepackage{color}
\usepackage{setspace}
\usepackage[round]{natbib}
\usepackage{amsmath}%
\usepackage{amsthm}%
\usepackage{palatino}
\usepackage{paralist}
\usepackage[top=8pc,bottom=8pc,left=8pc,right=8pc]{geometry}

\newtheorem{theorem}{Theorem}

\newtheorem{acknowledgement}[theorem]{Acknowledgement}

\newtheorem{definition}[theorem]{Definition}

\def\CM{\mathop{\rm CM}\nolimits}
\def\commentcolor{blue}
\newcount\commentnumber \commentnumber=0
\long\def\comment#1{\global\advance\commentnumber by 1 \bgroup\color{\commentcolor}Comment \number\commentnumber: #1\egroup}

\def\indep{\mathrel{\rlap{$\perp$}\kern1.6pt\mathord{\perp}}}
\newbox\dashbox
\setbox\dashbox=\vbox to 6pt{\vfill\hbox to 0.4cm{\hrulefill}\vfill}

\def\Section{\S}
\def\Sections{\S\S}

\def\W{{\cal W}}

\def\AI{\mathop{\rm AI}\nolimits}
\def\given{\mathrel{|}}

\def\vhalf{{\textstyle{\frac12}}}

\newif\ifignoretext \ignoretexttrue
\def\beginignoretext{\setbox0=\vbox\bgroup}
\def\endignoretext{\egroup \ifignoretext\relax\else\unvbox0 \fi}

\newbox\bigstrutbox
\setbox\bigstrutbox=\hbox{\vrule height 10.5pt depth3.5pt width 0pt}
\def\bigstrut{\relax\ifmmode\copy\bigstrutbox\else\unhcopy\bigstrutbox\fi}

\def\Real{{\mathbb R}}
\chardef\tie='176
\chardef\caret='136
\title{Statistical sparsity}

\author{Peter McCullagh\\
\textit{Department of Statistics}\\
\textit{University of Chicago}\thanks{Department of Statistics, University of Chicago,
5747 S. Ellis Ave, Chicago, Il 60637, U.S.A. E-mail:
pmcc@galton.uchicago.edu. Nick Polson is at ChicagoBooth, University of Chicago. E-mail: ngp@chicagobooth.edu.}\\
\\
Nicholas G. Polson\\
\textit{Booth School of Business}\\
\textit{University of Chicago}\\
\\}

\date{May 1, 2018
}

\begin{document}

\maketitle
\begin{abstract}
\noindent 
The main contribution of this paper is a mathematical definition of statistical sparsity, which is
expressed as a limiting property of a sequence of probability distributions.
The limit is characterized by an exceedance measure~$H$ and a rate parameter~$\rho > 0$,
both of which are unrelated to sample size.
The definition is sufficient to encompass all sparsity models that have
been suggested in the signal-detection literature.
Sparsity implies that $\rho$~is small, and a sparse approximation is 
asymptotic in the rate parameter, typically with error $o(\rho)$ in the sparse limit $\rho \to 0$.
To first order in sparsity, the sparse signal plus Gaussian noise convolution depends  on
the signal distribution only through its rate parameter and exceedance measure.
This is one of several asymptotic approximations implied by the definition, 
each of which is most conveniently expressed
in terms of the zeta-transformation of the exceedance measure.
One implication is that two sparse families having the same exceedance measure
are inferentially equivalent, and cannot be distinguished to first order.
A converse implication for methodological strategy is that it may be more fruitful to focus
on the exceedance measure, ignoring aspects of the signal distribution that have
negligible effect on observables and on inferences.
From this point of view, scale models and inverse-power measures seem particularly attractive.
\vspace{0.5pc}

\end{abstract}

\newpage
\section{Introduction}
\subsection{The role of a definition}
Statistical sparsity is concerned partly with phenomena that are rare,
partly with phenomena that are mostly zero,
but more broadly with phenomena that are  mostly negligible or 
seldom appreciably large.
Progress in mathematics is seldom impeded by inadequacy of definitions,
and the same may be said about progress in the development of sparsity as a 
concept in statistical work.
But, sooner or later, definitions are needed in order to clarify ideas and to keep confusion at bay.
The challenge is to formulate accurately a definition of sparsity that is 
faithful to current usage, and to explore its consequences.
Our approach uses a probabilistic limit.

Statistical sparsity is defined in \Section~2 as a limiting property of a sequence of 
probability distributions that governs both the rate at which probability 
accumulates near the origin, and the rate at which it decreases elsewhere.
Our definition covers all sparsity models that are found in the statistical literature
on sparse-signal detection and estimation.
It includes all two-group atom-and-slab mixtures, (Johnstone and Silverman 2004,  Efron, 2009),
all non-atomic spike-and-slab mixtures (George and McCulloch 1993;  Rockova and George 2018),
the low-index gamma model (Griffin and Brown, 2013),
and many Gaussian scale mixtures such as the
Cauchy scale family and the horseshoe scale family (Carvalho, Polson and Scott 2010).

The sparse limit is characterized by a rate parameter $\rho > 0$ and a measure~$H$ whose
product $\rho H$ determines the rarity of threshold exceedances.
The exceedance measure is also the chief determinant
of a certain restricted class of integrals, probabilities, and expected values
that arise in probabilistic assessments of signal activity.

In many cases, a definition tells us only what is intuitively well known.
But occasionally, a good mathematical definition reveals an aspect of the phenomenon that is
unexpected and not readily apparent from a litany of examples.
Sparsity is a case in point.
The phenomenon may be intuitively obvious, but the definition in terms of a
characteristic pair $(\rho, H)$ is much less so.
The first reason for a definition is that it highlights the role of the characteristic pair
and provides a definitive answer to the question of
whether a particular probability model is or is not statistically sparse,
in what way it deviates from sparsity, and so on.

The second reason is that the limit enables us to develop distributional
approximations for inferential purposes in sparse signal-detection problems,
i.e.,~approximations for the marginal distribution or the conditional distribution given the observation.
The sequence is essential because a sparse approximation is asymptotic in the rate parameter $\rho \to 0$,
and is unrelated to sample size and sample configuration.

The third reason is that, while the sequence of distributions determines the exceedance measure,
the exceedance measure does not determine the distributions.
Two sequences having the same exceedance measure are first-order equivalent in the sense that
all marginal and conditional distributions depend only on the exceedance measure.
For example, to certain atom-free spike-and-slab mixtures there corresponds an equivalent atom-and-slab mixture.
Likewise, the Cauchy and horseshoe scale families are equivalent, but they are
not equivalent to the low-index gamma model.

\subsection{Statistical implications}
The novelty of this paper lies entirely in the definition of sparsity,
which is statistically interesting on account of its implications.
We leave it to the reader to decide whether the implications described in \Sections~3--6 
are useful or relevant or have practical consequences,
but utility and practical considerations play no role in their derivation.
The over-riding implication is that it is futile to estimate any functional of the signal distribution
that is not first-order identifiable from the data that are observed.
Subsidiary implications flowing from the definition are as follows:
\begin{itemize}
\item the use of $(\rho, H)$ in place of the signal distribution for model specification;
\item the role of the asymptotic likelihood for parameter estimation (\Sections~3.4, \ref{illustrations});
\item the role of the zeta function for inference about the signal given the data (\Section~5);
\item the connection between scale models and inverse-power measures (\Section~4);
\item the interpretation of the Benjamini-Hochberg procedure in terms of conditional exceedance 
	rather than conditional false discovery or null signals (\Section~5.4).
\end{itemize}

The zeta transformation is defined in \Section~3;
it plays a key role for inference in the standard sparse signal detection model.
Section~4 focuses on the inverse-power exceedance measures,
a class that includes the sparse Cauchy model, the horseshoe model 
and all other scale families having similar tail behaviour.
Within the inverse-power class, there exists a particular family of probability distributions,
called the $\psi$-scale family, that has a highly unusual but extremely useful property.
Every Gaussian-$\psi$ convolution that arises in the signal-plus-noise model
is expressible exactly as a binary Gaussian-$\psi$ mixture.
This is a  closure or self-conjugacy property, which means that the observation
distribution belongs to the same family as the signal.

Section~\ref{conditionaldistribution} shows how the zeta function determines the asymptotic conditional distribution 
of the signal given the observation,
and Tweedie's formula for the conditional moment generating function.
The conditional activity, or $\epsilon$-exceedance probability, is shown to be a rational function of the
zeta transformation, which is closely related to the Benjamini-Hochberg procedure
(Benjamini and Hochberg 1995).

The theory is extended in \Section~\ref{hyperactivity} to a hyperactive random signal for which the 
asymptotic behaviour is technically more complicated.
Section~\ref{illustrations} illustrates the application of parametric maximum likelihood to estimate the
sparsity rate parameter and the exceedance index for subsequent inferential use.

\section{Sparse limit: definitions}
\subsection{Exceedance measure}
The sparse limit involves an exceedance measure, which is defined as follows.

\begin{definition}
A non-negative measure $H$ on the real line excluding the origin is termed an \emph{exceedance measure\/} if
$\int_{\Real\setminus\{0\}} \min(x^2, 1)\, H(dx) < \infty$.
A measure satisfying $\int_{\Real\setminus\{0\}} (1 - e^{-x^2\!/2}) H(dx) = 1$
is called a unit exceedance measure.
\end{definition}

Although the motivation for this definition is unconnected with stochastic processes,
every exceedance measure is the L\'evy measure of an infinitely divisible distribution
on the real line, and vice-versa.
No constraint is imposed on the total measure, which may be finite or infinite.

To every non-zero exceedance measure $H$ there corresponds a ray 
$\{\lambda H(dx) : \lambda > 0\}$ of proportional measures.
Each ray contains as a reference point a unit measure such that
$(1 - e^{-x^2\!/2}) H(dx)$ is a probability distribution on $\Real{\setminus}\{0\}$.
For example, the unit inverse-power measures are
\begin{equation}\label{inversepowerexceedance}
H(dx) = \frac{d\, 2^{d/2 - 1}}{ \Gamma(1-d/2)} \frac{dx}{|x|^{d+1}}
\end{equation}
for $0 < d < 2$.
\begin{definition}
The activity index $0\le\AI(H) < 2$ gauges the behaviour in a neighbourhood of the origin:
\[
\AI(H) = \inf \bigl\{\alpha > 0 : \int_{-1}^1 |x|^\alpha \, H(dx) < \infty \bigr\}.
\]
Every finite measure has activity index zero; the measure (\ref{inversepowerexceedance}) has activity index~$d$.
\end{definition}

\comment{The activity index $\AI(H)$ is strictly less than two because continuity of
$\alpha \mapsto  \int_{-1}^1 |x|^\alpha \, H(dx)$ for $\alpha > 0$ implies that
$\{\alpha>0 : \int_{-1}^1 |x|^\alpha \, H(dx) < \infty \}$ is an open set containing~2.
In particular, the limit $\lim_{\epsilon\to 0} \epsilon^{2-\AI(H)} =0$ arises in \Section~\ref{zetainequalities}.}

\begin{definition}
The space $\W^\sharp$ of L\'evy-integrable functions
consists of bounded continuous functions $w(x)$ on the real line such that
$x^{-2} w(x)$ is also bounded and continuous.
L\'evy integrability implies  $\int_{\Real\setminus\{0\}} w(x)\, H(dx) < \infty$ for 
every $w\in\W^\sharp$ and every exceedance measure~$H$.
\end{definition}

The functions $\min(x^2, 1)$, \ $x^2 e^{-x^2}$ and  $1 - e^{-x^2\!/2}$ belong to~$\W^\sharp$.

\subsection{Sparse limit}
Let $\{P_\nu\}$ be a sequence of probability distributions indexed by $\nu > 0$,
and converging weakly to the Dirac measure $\delta_0$ as $\nu \to 0$.
Sparsity is a rate condition governing the approach to the weak limit.

\begin{definition} 
A sequence of probability distributions $\{P_\nu\}$  is said to have a sparse limit with rate~$\rho_\nu$
if there exists a unit exceedance measure $H$ such that
\begin{equation}\label{exceedancelimit}
\lim_{\nu \to 0} \rho_\nu^{-1} \int_\Real w(x)\, P_\nu(dx) = \int_{\Real\setminus\{0\}} w(x)\, H(dx)
\end{equation}
for every $w\in\W^\sharp$.
Otherwise, if the limit is zero for every~$w$, the sequence is said to be sparse with rate $o(\rho_\nu)$.
\end{definition}

The motivation for this definition comes from extreme-value theory, which focuses on 
exceedances over high thresholds
(Davison and Smith 1990).
Each sparse-signal threshold $\epsilon > 0$ is fixed as $\nu \to 0$, but is automatically high relative to the bulk of the distribution:
in this respect, the parallel with extreme-value theory is close.
Unlike extreme-value theory, sparsity places no emphasis on
limit distributions for the excesses over any threshold.
Formally setting $w(\cdot)$ equal to the indicator function for the
event $\epsilon^+ = (\epsilon, \infty)$ or $[\epsilon, \infty)$ in the integrals (\ref{exceedancelimit}) 
gives the  motivating condition---that the sparsity rate is the rarity of exceedances
\begin{equation}\label{exceedancedef}
\lim_{\nu \to 0} \rho_\nu^{-1} \,P_\nu(\epsilon^+) = H(\epsilon^+) < \infty.
\end{equation}
There is a similar limit for negative exceedances, and any other subset whose closure does not include zero.

\comment{The integral definition implies (\ref{exceedancedef}), 
but the converse fails if the limit in (\ref{exceedancedef}) is not a L\'evy measure.
For example, the Dirac-Gaussian mixture $(1-\nu)\delta_0 + \nu N(0, \nu^{-1})$ does not have a sparse limit,
but (\ref{exceedancedef}) is satisfied by $\rho_\nu = \nu$ with $H(\epsilon^+) = 1/2$.
Section~\ref{hyperactivity} discusses another example where the limit measure is non-trivial
but not in the L\'evy class.}

Since the definition involves only the limit  $\nu\to 0$, it is always possible to
re-paramet\-erize by the rate function, so that $\nu = \rho$.
This standard parameterization is assumed where it is convenient.

\begin{definition} {\rm Sparse-limit equivalence:}
Regardless of their parameterization, two sparse families having the same exceedance measure
are said to be equivalent in the sparse limit.
\end{definition}

Let $\{P_\nu\}$ and $\{Q_\nu\}$ be two families having the same unit exceedance measure~$H$,
both taken in the standard parameterization with rate parameter $\rho = \nu$.
In effect, the rate parameterization matches each distribution in one family with a sparsity-matching
distribution in the other, so the families are in 1--1 correspondence, at least in the approach to the limit.
For any function $w\in \W^\sharp$, the limit integrals are finite and equal:
$$
\lim_{\nu \to 0} \nu^{-1}\int w(x) \, P_\nu(dx) = \lim_{\nu\to 0} \nu^{-1} \int w(x)\, Q_\nu(dx)
        = \int w(x)\, H(dx)
$$
Consequently, near the sparse limit, both integrals may be approximated by
$$
\int w(x)\, P_\nu(dx) \simeq \int w(x)\, Q_\nu(dx) = \nu \int w(x)\, H(dx) + o(\nu).
$$
This analysis implies that every $\W^\sharp$-integral using $P_\nu$ 
as the signal distribution is effectively the same as the integral using $Q_\nu$
at the corresponding sparsity level.

\begin{definition}{\rm Sparse scale family:}
A scale family of distributions with density $\sigma^{-1} p(x/\sigma)$ is called a sparse scale family
if it is sparse in the small-scale limit~$\sigma \to 0$.
The rate function $\rho(\sigma)$ need not coincide with the scale parameter.
\end{definition}

The Student-$t$ scale family on $d < 2$~degrees of freedom is a sparse scale family
with rate function $\rho(\sigma) \propto \sigma^{d}$ and exceedance density proportional to~$dx/|x|^{d+1}$.

If the scale family $\{P_\sigma\}$ is sparse with rate parameter $\rho(\sigma)$, then, for small~$\sigma$,
$$
P_\sigma(dx) = p(x/\sigma)\, dx/\sigma \simeq \rho(\sigma) \, h(x)\, dx.
$$
Setting $x=1$ and $u=1/\sigma$ gives $p(u) \simeq h(1) \rho(u^{-1})/u$ as $u\to \infty$.
Conversely, $h(x) = x^{-1} \rho(\sigma/x)/\rho(\sigma)$ for $x > 0$, implying that
 $\rho(\sigma) = \sigma^d$ for some power $d > 0$.
If $p$ is not symmetric, the power index for $x < 0$ may be a different number.
It follows that the exceedance density of a sparse scale family
is an inverse power function $h(x) \propto 1/x^{d+1}$, the same as the tail behaviour
of $p(x)$ as $x \to \infty$.

%
%
%
\subsection{Infinite divisibility}\label{infinitedivisibility}
Let $F$ be an infinitely divisible probability distribution on the real line, 
and let $\{F_\nu\}$ be the L\'evy family
indexed by the convolution parameter, i.e.,~$F_\nu \star F_{\nu'} = F_{\nu+\nu'}$ with $F_1 = F$.
The L\'evy process is sparse with rate~$\nu$, and the exceedance measure is the L\'evy measure
(Barndorff-Nielsen and Hubalek, 2008).

To each exceedance measure there corresponds an infinite equivalence class of sparse sequences,
most of which are not closed under convolution.
This result tells us that each equivalence class contains exactly one L\'evy process;
the zero equivalence class contains the Gaussian family, i.e.,~the Brownian motion process.
Despite this characterization, exceedance measures and L\'evy processes have not played 
a prominent role in either frequentist or non-frequentist work on sparsity.
 

\comment{A typical spike-and-slab distribution is not infinitely divisible.
However, there are exceptions.
For each positive pair $(\lambda, \tau)$, the atom-and-slab lasso distribution
\[
e^{-\lambda}\delta_0(x) + (1-e^{-\lambda}) \tau e^{-\tau |x|}/2
\]
is infinitely divisible with finite L\'evy measure 
\begin{equation}\label{laplacelevy}
H_{\lambda,\tau}(dx) \propto \Bigl( e^{-\tau|x|} - e^{-\tau e^{\lambda/2}|x|} \Bigr) |x|^{-1}\, dx.
\end{equation}
For each $(\lambda, \tau)$, the family $\{F_\nu\}$ exists, but the distributions are not easily exhibited.

The atom-and-slab lasso family with $\rho=1 - e^{-\lambda}$ as the sparsity rate parameter is
not to be confused with the L\'evy family $\{F_\nu\}$ in which $\lambda>0$ is held fixed.
The  exceedance measures are $H_{\lambda,\tau}$ and $\tau e^{-\tau|x|}/2 = \lim_{\lambda\to 0} \lambda^{-1} H_{\lambda,\tau}$.
}

\subsection{Sparsity expansion}
Weak convergence of the sequence $\{P_\nu\}$ to the Dirac measure $\delta_0$ is concerned 
with the behaviour of $P_\nu$-integrals for a suitable class of functions.
For any bounded continuous function~$w$ having one  continuous derivative at zero, the symmetrized function
\[
2\tilde w(x) = w(x) + w(-x) - 2w(0)
\]
is $O(x^2)$ near the origin.
Thus $\tilde w \in \W^\sharp$ is L\'evy integrable
and, if $P_\nu$ is symmetric, sparseness implies a linear expansion for small~$\nu$:
\begin{eqnarray*}
\int_\Real w(x)\, P_\nu(dx) &=& w(0) + \int_\Real \bigl(w(x) - w(0)\bigr) \, P_\nu(dx) \\
	&=&  w(0) +  \int_\Real \tilde w(x) \, P_\nu(dx) \\
	&=&  w(0) + \rho \int_{\Real\setminus\{0\}} \tilde w(x) \, H(dx) + o(\rho).
\end{eqnarray*}
The exceedance measure is the directional derivative or linear operator governing
the approach to the weak limit:
\[
P_\nu(w) - \delta_0(w) = \rho H(\tilde w) + o(\rho)
\]
in the sense of integrals.
Sparsity determines the difference $P_\nu(w) - \delta_0(w)$,
but only to first order in~$\rho$.

\subsection{Examples}\label{Examples}
For $\epsilon > 0$, the exceedance event $A_\epsilon\subset\Real$, or activity event, is
the complement of the closed interval $\bar A_\epsilon = [-\epsilon, \epsilon]$.

\smallskip
Example 1: The $\epsilon$-exceedance probability for the Laplace distribution with density 
$\sigma^{-1} e^{-|x|/\sigma} / 2$  is $e^{-\epsilon/\sigma}$, implying
$\lim_{\sigma\to 0} \sigma^{-p} e^{-\epsilon/\sigma} = 0$ for every $\epsilon > 0$ and $p>0$.
The scale family is sparse with rate $o(\sigma^p)$ for every $p > 0$.
There is no definite sparsity rate parameter satisfying (\ref{exceedancelimit}) with a finite non-zero limit,
so we say that the sequence belongs to the zero-activity class.
The Gaussian scale family, and all other scale families having exponential tails,
have the same property.
So far as this paper is concerned all sparse families in this class are trivial and equivalent.

\smallskip
Example 2: Let $F$ be a probability distribution on the real line.
The atom and slab family $P_\nu = (1-\nu)\delta_0 + \nu F$ indexed by
$0 < \nu \le 1$ is sparse with exceedance measure proportional to~$F$.
The unit exceedance measure is $F/K$, where $K=\int(1-e^{-x^2/2})\, F(dx)$,
and the exceedance rate is $\rho = K\nu$, so the product satisfies $\rho H = \nu F$.
The activity-reduction factors $\rho/\nu$ for the standard Laplace, Cauchy and
Gaussian distributions are $0.34$, $0.48$ and $0.29$ respectively.

The mixture-indexed spike and $F$-slab family with a fixed scale parameter
belongs to the finite class;
it is not to be confused with the atom-free $F$-scale family or intermediate combinations.

\smallskip
Example 3a: The family of double gamma distributions with density
$$
p_\nu(x) = \frac{|x|^{\nu - 1} e^{-|x|}} {2\, \Gamma(\nu)}
$$
is sparse with rate parameter $\rho \propto \nu$ and exceedance density
$h(x) \propto |x|^{-1} e^{-|x|}/2$.
The total mass $\lim_{\epsilon \to 0} H(A_\epsilon)$ is infinite but there is no atom at zero.

\smallskip
Example 3b: For fixed $\sigma$, the re-scaled double gamma family with density 
$\sigma^{-1} p_\nu(x/\sigma)$
is sparse with rate parameter $\rho \propto \nu$ and exceedance density
$h(x) \propto  |x|^{-1} e^{-|x|/\sigma}$,
which depends on~$\sigma$.
Each of the sub-families for different~$\sigma > 0$ has its own exceedance measure.
These are not equivalent because they are not proportional.

For fixed index~$\nu$, the double gamma scale family is also sparse
with rate $o(\sigma^p)$ for every $p > 0$,
so the scale family belongs to the zero-activity class.

\smallskip
Example 4: The Cauchy family $C(\sigma)$ with probable error $\sigma > 0$ and density
$$
P_\sigma(dx) = \frac{\sigma\, dx} {\pi ( \sigma^2 + x^2)}
$$
is sparse with inverse-square exceedance $dx/(\sqrt{2\pi}\, x^2)$
and rate  $\rho = \sigma\sqrt{2/\pi}$.
The  scale families
\[
C(\sigma),\quad \vhalf\bigl(\delta_0 + C(2\sigma)\bigr), \quad 0.75\, N(0, \sigma^2) + 0.25\, C(4\sigma),
\]
have the same rate parameter and exceedance measure.
Similar remarks apply to a large number of families that have been proposed
as prior distributions in the literature,
including the scale family generated by various Gaussian mixtures such as the horseshoe 
distribution with density $\log(1+1/x^2) / (2\pi)$.

\smallskip
Example 5a: Consider the distribution with density
$$
p(x) = \frac{1 - e^{-x^2/2}} { x^2\, \sqrt{2 \pi}}
$$
and let $p_\sigma(x) = \sigma^{-1} p(x/\sigma)$ be the density of the re-scaled distribution.
Then the family $\{P_\sigma\}$ is sparse with rate parameter $\rho = \sigma$
and inverse-square exceedance.

\smallskip
Example 5b:
For $d > 1/2$, let $w(x) = x^{2d}/(1+x^2)^d$, and let
$$
p_\sigma(x) = \frac{\Gamma(d)} {\Gamma(d-1/2) \surd\pi} \frac{\sigma\, w(x/\sigma )} {x^2}.
$$
This scale family is sparse with the same rate function and inverse-square exceedance measure as the previous two.
The weight function has no role in the limit provided that
the integral is finite, $w(x) \sim x^2$ for small~$x$, and $w(x) \to 1$ as $x\to \pm\infty$.

\smallskip
Example 6:
The Dirac-Gaussian mixture $P_\nu = (1-\nu)\delta_0 + \nu N(0, 1/\nu)$
converges to a point mass, as does $(1-\nu)\delta_0 + \nu N(0, \nu)$,
but neither mixture has a sparse limit according to the definition.
The Dirac-Cauchy mixture  $(1-\nu)\delta_0 + \nu C(\nu)$ has a sparse inverse-square limit
with rate parameter $\rho = \nu^2$, but  $(1-\nu)\delta_0 + \nu C(1/\nu)$ does not.

\section{Zeta function and zeta measure}
\subsection{Definitions}\label{zetadefinition}
We assume henceforth that every sparse family is symmetric, and the exceedance measure is
expressed in unit form, so that $\int(1 - e^{-x^2/2}) H(dx) = 1$.

To each exceedance measure $H$ there corresponds a \emph{zeta function}
\begin{eqnarray}
\label{zetadef}
\zeta(t) &=& \int_{\Real\setminus\{0\}} \bigl(\cosh(tu) - 1\bigr) e^{-u^2\!/2} \,H(du),
\end{eqnarray}
which is positive and finite, symmetric and convex, satisfying~$\zeta(0) = 0$.
By construction, the zeta function is the cumulant function of the infinitely divisible distribution
with down-weighted L\'evy measure $e^{-u^2\!/2} H(du)$.
The zeta function is analytic at the origin, so this L\'evy process has finite moments of all orders.

Numerical values are shown in Table~1 for three inverse-power measures.

The \emph{zeta measure} is the integrand in (\ref{zetadef}):
\begin{equation}\label{zetadistribution}
\zeta(du; \theta) = \bigl(\cosh(\theta u) - 1) e^{-u^2\!/2} \,H(du),
\end{equation}
which is a weighted linear combination of symmetric measures $|u|^{2r} e^{-u^2/2}H(du)$
with coefficients $\theta^{2r}\!/(2r)!$ for $r\ge 1$, and finite total mass~$\zeta(\theta)$.
The zeta measure occurs as one of two components of the conditional distribution in \Section~\ref{symmetrization}.

\begin{table}
\small
\begin{tabular}{crrrrrrrrrrrrr}
\multicolumn{14}{l} {\hbox{Table 1: Zeta function  $\zeta_d(x)$ for three inverse-power exceedance measures.}}\\
\hline
\noalign{\smallskip}
$\ d\,\backslash \, x$ &2.0&2.2&2.4&2.6&2.8&3.0&3.2&3.4&3.6&3.8&4.0&4.2&4.4\\
\hline
\noalign{\smallskip}
0.5 &1.9   &2.7   &3.9   &5.8   &8.8  &13.9  &22.9  &39.4  &70.9 &133.7 &264.3 &547.6 &1188.4 \\
1.0   &3.1   &4.2   &5.8   &8.1  &11.6  &17.2  &26.5  &42.7  &72.5 &129.6 &244.2 &485.0 &1013.9\\
1.5  &3.8   &4.8   &6.2   &8.1  &10.7  &14.4  &20.2  &29.5  &45.5  &74.4 &129.8 &241.6 &478.7\\
\hline
\end{tabular}
\end{table}

\subsection{Tail inflation factor}
The zeta function is an integral transformation much like a Laplace transform,
i.e.,~formally $H$~is a measure on the observation space and $\zeta$~is a function on the dual space.
However, if $\phi(x)$ is the standard normal density,
the product  $\psi(x) = \phi(x) \zeta(x)$ is also a probability density with characteristic function
\begin{eqnarray}
\nonumber
\int e^{i t x} \psi(x)\, dx &=& \int_{\Real\setminus\{0\}} e^{-u^2\!/2} \,H(du)
	 \int_\Real \phi(x) \bigl(e^{x(u+it)}\!/2 + e^{-x(u-it)}\!/2 - e^{i t x})\, dx, \\
\nonumber
	 &=& e^{-t^2/2} \int_{\Real\setminus\{0\}} \bigl(\cos(t u) - e^{-u^2/2} \bigr) \, H(du), \\
\label{psicf}
	&=& e^{-t^2/2} - e^{-t^2/2} \int_{\Real\setminus\{0\}} \bigl(1 - \cos(t u) \bigr) \, H(du) .
\end{eqnarray}
Provided that $H$ is a unit exceedance measure, the value at $t=0$ is one.

Section~\ref{section:marginaldistribution} shows that~$\psi$ is the tail-inflation component 
of the marginal distribution of the observations.
The left panel of Figure~1 shows the density function for the inverse-power exceedance measures,
all of which have similar bimodal distributions differing chiefly in modal height and tail behaviour.
There is a certain qualitative similarity with a pair of distributions depicted in Figure~1 of
Johnson and Rossell (2009),
and recommended there as priors for testing a Gaussian mean.

%
For typical exceedance measures having regularly-varying tails, 
the characteristic function of $\psi$ is not analytic at the origin, 
in which case the distribution does not have finite moments.
For example, if $H$ is 
the inverse-power exceedance (\ref{inversepowerexceedance}) for some $0 < d < 2$,
the characteristic function (\ref{psicf}) reduces to
\begin{equation}\label{inversepowercf}
\int_\Real e^{i t x} \psi(x)\, dx = e^{-t^2/2} ( 1 - |t|^d / K_d ),
\end{equation}
where $K_d = 2^{d/2} \Gamma(1/2 + d/2)/\sqrt{\pi}$.
By contrast, if $H(du) = K e^{-\alpha|u|} du/|u|$ has exponential tails, 
the integral in (\ref{psicf}) is
\[
\int_{\Real\setminus\{0\}}\bigl(1 - \cos(tu)\bigr)\, H(du) =  K \log(1+t^2/\alpha^2) ,
\]
which is analytic at the origin.
A derivation is sketched near the end of \Section~\ref{section:convolutionmixture}.

\subsection{Sparsity integrals}
To see how the zeta function arises in sparsity calculations, consider first the integral of the weighted distribution
\begin{eqnarray*}
\int e^{-x^2/2}\, P_\nu(dx) &=& 1 - \int (1 - e^{-x^2/2})\, P_\nu(dx),\\
	&=& 1 - \rho \int(1 - e^{-x^2/2})\, H(dx) + o(\rho),\\
	&=& 1-\rho + o(\rho),
\end{eqnarray*}
where $\rho \equiv \rho(\nu)$ is the rate function.
Second, for any sparse family, the Laplace transform is
\begin{eqnarray*}
\int_\Real e^{t x} e^{-x^2/2}\, P_\nu(dx) 
	&=& 1 - \rho + \int_\Real (e^{ t x} - 1) e^{-x^2/2} \, P_\nu(dx) \\
	&=& 1 - \rho + \int_\Real (\cosh(tx) - 1) e^{-x^2/2} \, P_\nu(dx),\\
	&=& 1 - \rho + \rho \int_{\Real\setminus\{0\}} (\cosh(tx) - 1)\, e^{-x^2/2} H(dx) + o(\rho), \\
	&=& 1-  \rho + \rho \zeta(t) + o(\rho).
\end{eqnarray*}
The normalized family $(1-\rho)^{-1} e^{-x^2/2} P_\nu$ is sparse with rate~$\rho'=\rho/(1-\rho)$, 
exceedance measure $e^{-x^2/2} H(dx)$, and Laplace transform $1+\rho' \zeta(t) + o(\rho)$.

\subsection{Signal plus noise convolution}\label{section:marginaldistribution}
Suppose that the observation $Y$ is a sum of two independent unobserved random variables
$Y = \mu + \varepsilon$, where the signal $\mu \sim P_\nu$ is sparse, and $\varepsilon \sim N(0, 1)$
is a standard Gaussian variable.
Then the joint density of $(Y, \mu)$ at $(y, u)$ is $\phi(y-u) \,p_\nu(u)$, while the marginal density of
the observation is a Gaussian-$\psi$ mixture:
\begin{eqnarray}
\nonumber
m_\nu(y) &=& \int_\Real \phi(y - u) \, P_\nu(du) \\
\nonumber	&=&  \phi(y) \int_\Real e^{y u - u^2/2} \, P_\nu(du) \\
\nonumber	&=& \phi(y)\bigl(1 - \rho + \rho \zeta(y) \bigr)  +o(\rho)\\
\label{marginalmixture}	&=& (1-\rho) \phi(y) + \rho \psi(y) + o(\rho).
\end{eqnarray}
For every sparse family, the sparsity parameter satisfies $1-\rho = m_\nu(0)/\phi(0)$, 
which provides a recipe for consistent estimation.
Alternatively, if $H$, $\zeta$ or $\psi$ are given, the mixture (\ref{marginalmixture}) can be
 fitted directly by maximum likelihood: see \Section~\ref{illustrations}.

\comment{The Gaussian-$\psi$ mixture is a consequence of sparsity alone.
It is compatible with a sparse atom-and-slab mixture $(1-\nu)\delta_0 + \nu F$ for signals,
but the sparsity rate is strictly less than the slab weight
$\rho/\nu = \int (1-e^{-u^2/2}) F(du) < 1$.
The identity $\psi(0) = 0$ implies that $\psi$~cannot be the response distribution for any signal subset,
so the sparsity rate in the marginal mixture (\ref{marginalmixture}) is not interpretable as the
prior probability of a non-null signal in the standard two-group model
(Johnstone and Silverman, 2004;
Efron, 2008, 2010).

Although $(\rho, H)$ determines the marginal density to first order,
Example~4 in \Section~\ref{Examples} show that $(\rho, H)$ 
does not determine the null fraction $0 \le P_\nu(\mu=0) < 1-\rho$, which may be zero.}

\begin{figure}
\centerline{
\includegraphics[scale = 0.27]{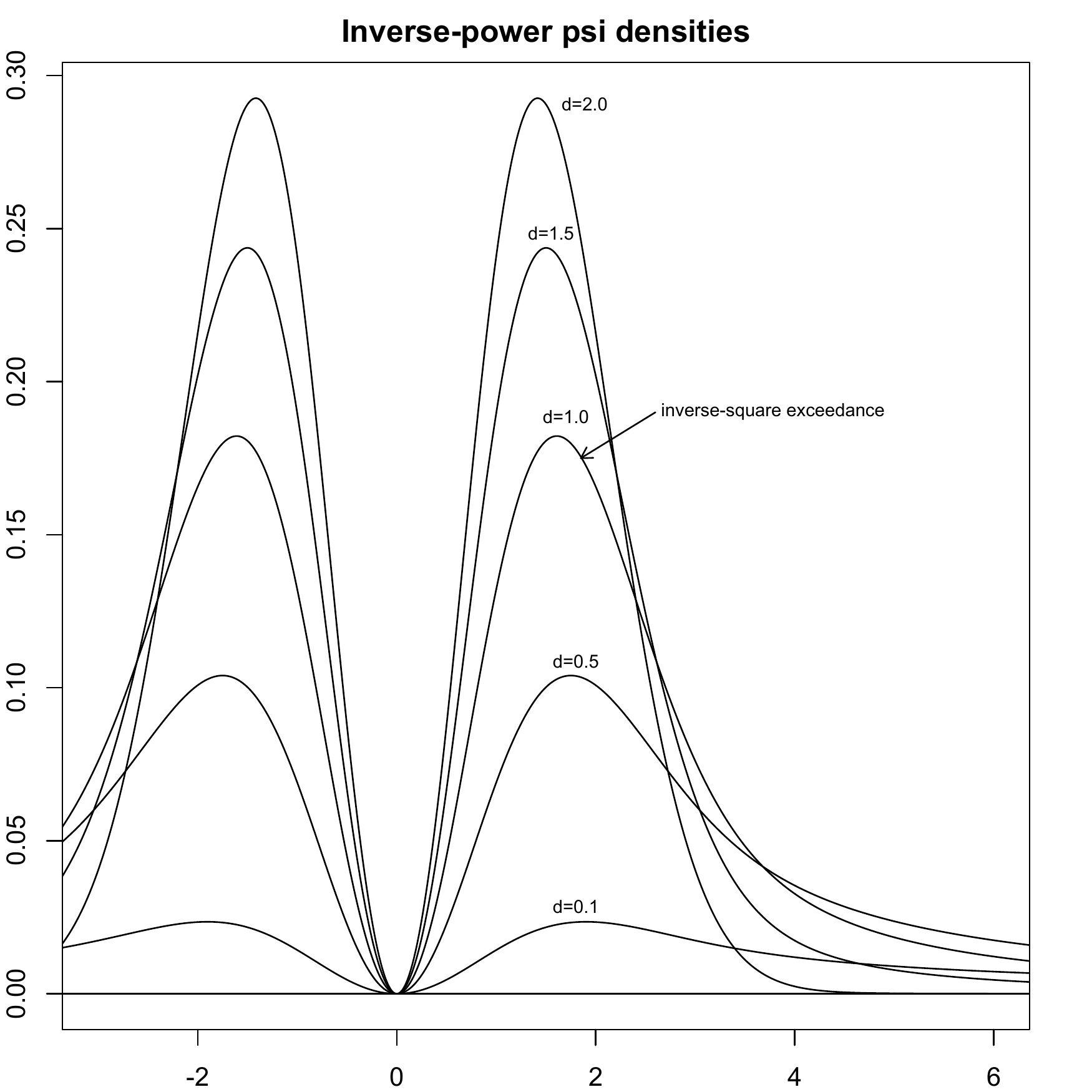}\hfil
\includegraphics[scale = 0.27]{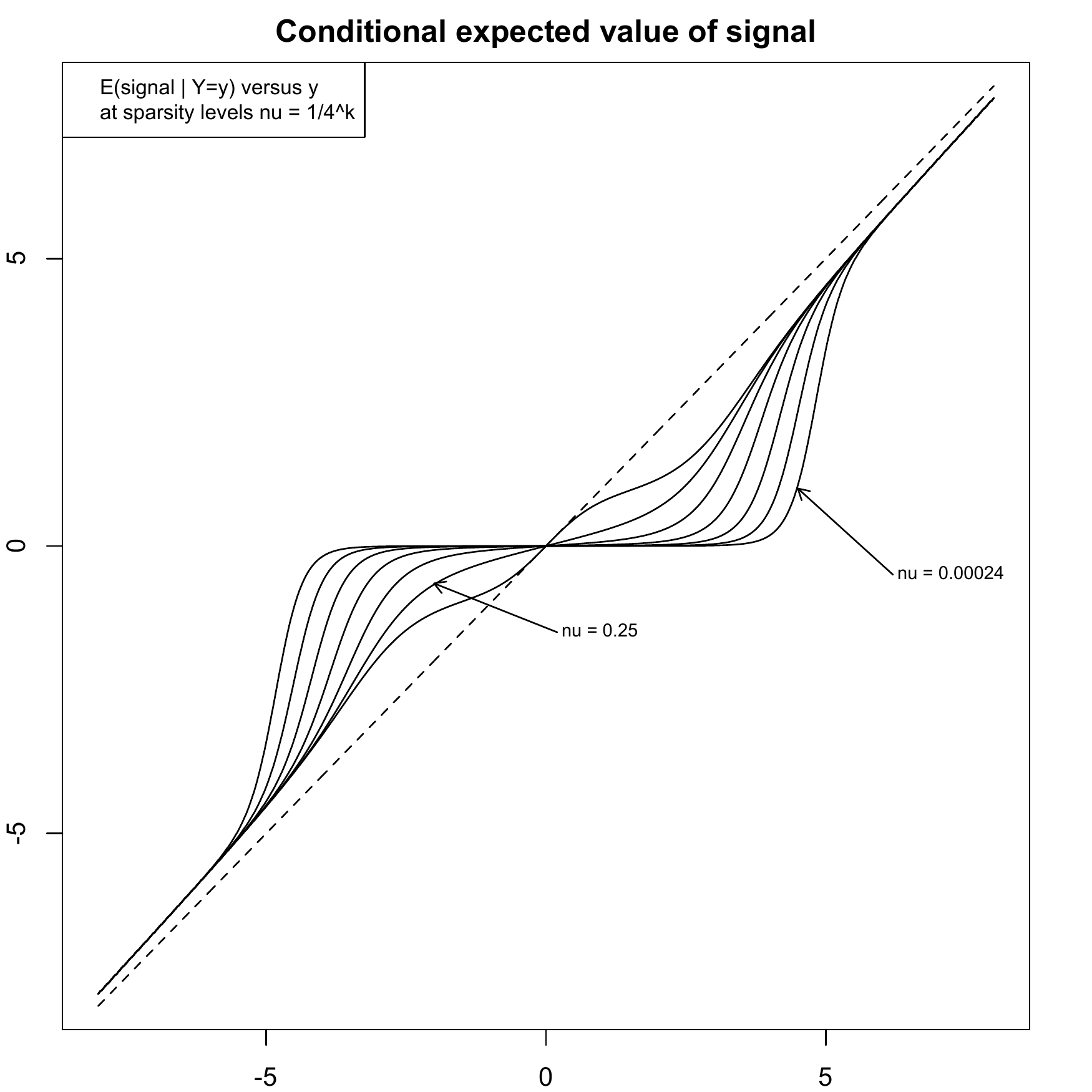}\hfil
\includegraphics[scale = 0.27]{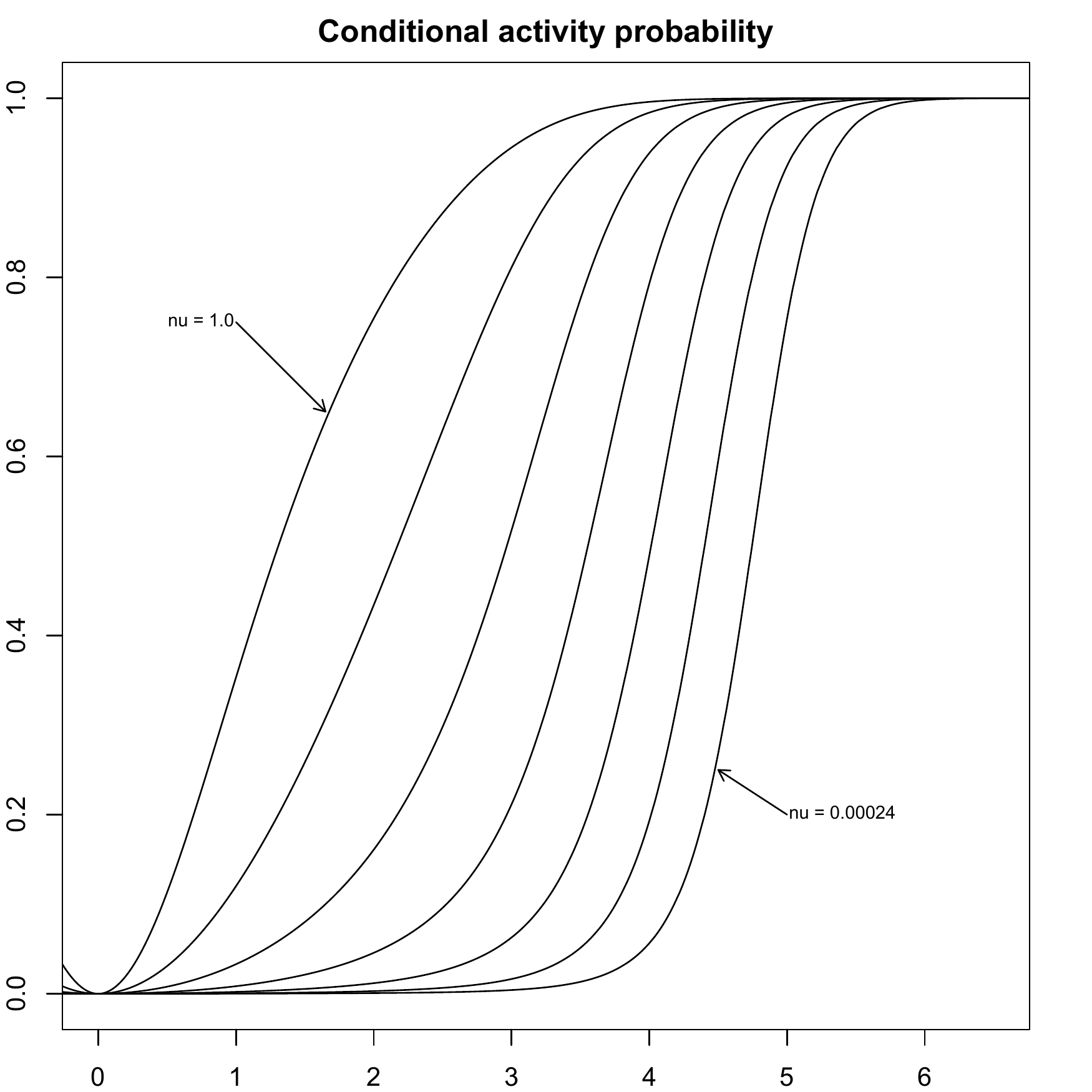}\hfil
}
\caption{Left panel: tail inflation densities $\psi(x)$ for the inverse power exceedance measures.
Middle panel: Bayes estimate of the signal using the inverse-square exceedance measure;
Right panel: conditional activity probability (\ref{exceedanceprob}) as a function of~$y$ for a range of sparsity values $4^{-k}$, $0 \le k \le 6$.}
\end{figure}

\subsection{Two inequalities}\label{zetainequalities}
For fixed~$t$, the zeta function is linear in~$H$, so the maximum necessarily occurs
on the boundary at an atomic measure $H(\{\pm u\}) = 1/(1-e^{-u^2/2})$ for some $u \neq 0$, 
or, if the maximum does not exist, the supremum occurs in the limit $u \to 0$\null. 
For $t^2 \le 3$, the limit point prevails, and the supremum is 
\[
\zeta(t) \le \lim_{u \to 0} \frac{\cosh(t u) - 1} {e^{u^2/2} - 1} = t^2,
\]
which is attained in the limit $d\to 2$ for the inverse-power class.
Accordingly, the leading Taylor coefficient  $\zeta_2 = \int u^2 e^{-u^2\!/2} H(du)$ satisfies $\zeta_2 < 2$.

The second inequality is concerned with the behaviour of the zeta measure in a neighbourhood of the origin.
If $H$~has an atom at $u\neq 0$, then $\zeta$~also has an atom at~$u$.
But $\AI(H) < 2$ implies that there is no atom at the origin,~i.e., for each $\theta$
\begin{equation}\label{nozeroatom}
\lim_{\epsilon \to 0} \zeta((-\epsilon, \epsilon); \theta) = \frac{K \theta^2}{2-\AI(H)} \lim_{\epsilon\to 0}  \epsilon^{2-\AI(H)} = 0.
\end{equation}
It follows that the density limit $\lim_{\epsilon\to 0}\epsilon^{-1}\zeta((-\epsilon, \epsilon); \theta)$ is 
zero for low-activity measures $\AI(H) < 1$ and infinite for $\AI(H) > 1$: see~Fig.~3.

For $|u| \le |u'|$, the series expansion with positive coefficients implies 
\[
\frac{\cosh(u) - 1} {u^2} \le \frac{\cosh(u')-1} {u^{\prime 2}}.
\]
For every $\theta$ and $\epsilon > 0$, it follows that
\begin{eqnarray}
\nonumber
\zeta\bigl((-\epsilon, \epsilon); \theta\bigr) 
	&=& \theta^2 \int_{-\epsilon}^\epsilon \frac{(\cosh(\theta u) - 1)}{\theta^2 u^2} u^2 e^{-u^2/2} \, H(du)\\
\nonumber
	& \le& \frac{\cosh(\theta \epsilon) - 1} {\epsilon^2} \int _{-\epsilon}^\epsilon u^2 e^{-u^2/2}\, H(du) \\
\label{zetaupperbound}
	& \le & \zeta_2\, \frac{\cosh(\theta \epsilon) - 1} {\epsilon^2}.
\end{eqnarray}
For example, $\zeta_2 < 2$ implies $\zeta\bigl((-1, 1); \theta\bigr) < 2\cosh(\theta) - 2$.

\section{Inverse power exceedances}\label{ipex}
\subsection{Summary}
This section summarizes, without proofs, the role of the zeta function~(\ref{zetadef})
for sparse scale families whose unit exceedance densities
for $0 < d < 2$ are shown in (\ref{inversepowerexceedance}).
It has the following properties.
\begin{enumerate}
\item The zeta function is expressible as a power series
\begin{equation}\label{zetad}
\zeta(x) = \frac{d(2-d)}{\Gamma(2-d/2)} \sum_{r=1}^\infty \frac{2^{r-2}\, \Gamma(r - d/2)\, x^{2r}} {(2r)!} ,
\end{equation}
which is symmetric with infinite radius of convergence.
\item The characteristic function of $\psi$ is $e^{-t^2/2}\bigl(1 - |t|^d/K_d\bigr)$, 
	where $K_d$ is given in~(\ref{inversepowercf}).
\item The scale family $\sigma^{-1} \psi(y/\sigma)$ is sparse with rate parameter $\rho = \sigma^{d}$
	and the same exceedance density~(\ref{inversepowerexceedance}).
\item Let $\eta \sim \psi$ and $\epsilon\sim \phi$ be independent.  To each pair of coefficients with
	 $a^2 + b^2 = 1$ there corresponds a number $0 \le \alpha \le 1$ such that the
	linear combination $a \epsilon + b \eta$ is distributed as the mixture 
	$(1-\alpha) \phi + \alpha \psi$.
\item The marginal distribution of $Y$ is a mixture $m_\nu(y) = (1-\rho) \phi(y) + \rho \psi(y)$.
\end{enumerate}

The statement in 5 is a little imprecise because the signal distribution is not mentioned.
Modulo the scale factor $\sqrt{1+\sigma^2}$ that occurs implicitly in~4, 
the statement is exact if the signal is distributed as $\sigma^{-1}\psi(\cdot/\sigma)$ for
arbitrary~$\sigma$, which determines the mixture weight~$\rho$.
It is also correct, modulo a similar scale factor, if the signal is distributed according to the mixture itself.
More importantly, it is correct for every signal distribution in this inverse-power equivalence class,
but with error $o(\rho)$ in the approach to the sparse limit.

\subsection{The convolution-mixture theorem}\label{section:convolutionmixture}
With $\zeta$ defined by its series expansion (\ref{zetad}), the probability distribution with density
\[
\psi(x) = \phi(x) \zeta(x)
\]
has a remarkable property that makes it singularly well adapted for statistical applications related to 
sparse signals contaminated with additive Gaussian error.
Not only is the $\psi$-scale family sparse with inverse-power exceedance measure, but
each Gaussian-$\psi$ convolution is also expressible as a binary Gaussian-$\psi$ mixture as follows.

\begin{theorem}
Convolution-mixture:  For arbitrary scalars $a, b$, let  
\[
Y = a \varepsilon + b \eta,
\]
where $\varepsilon\sim N(0,1)$ and $\eta \sim \psi$ are independent random variables.
Also, let $Y'$ be the mixture
\[
Y' = \left\{ \begin{array}{cl}
	\eta\,\sqrt{a^2 + b^2} &\strut \hbox{ with probability } \alpha = |b|^d/ (a^2+b^2)^{d/2}  \\
	\noalign{\smallskip}
	\varepsilon\,\sqrt{a^2+b^2} &\strut \hbox{  otherwise.}
	\end{array}
	\right.
\]
Then  $Y$ and $Y'$ have the same distribution denoted by $\CM_d(\alpha, a^2+b^2)$.
\end{theorem}

The theorem states that every linear combination with norm $\sigma=\sqrt{a^2+b^2}$ is equal in distribution
to an equi-norm Gaussian-$\psi$ mixture.
An arbitrary binary mixture is not expressible as a convolution unless the two scale parameters are equal.

\begin{proof}
The result is a direct consequence of the characteristic function (\ref{inversepowercf}).
One derivation proceeds from (\ref{psicf})  by analytic continuation of the gamma integral:
\[
\int_0^\infty u^{\alpha-1} e^{-s u} (2 - e^{- i t u} - e^{i t u}) \, du = 
	\Gamma(\alpha) \biggl( \frac 2 {s^\alpha} - \frac1 {(s+it)^\alpha} - \frac1 {(s-it)^\alpha} \biggr),
\]
which is convergent for $s > 0$ and $\alpha > -2$.
For $-2 < \alpha< 0$, the limit as $s \to 0$ is 
\[
\int_{\Real\setminus\{0\}} \frac{ (1 - \cos(tu)) \, du }{ |u|^{d+1}} = \frac{2\cos(d\pi/2)\Gamma(2-d)} {d(1-d)} |t|^d,
\]
where $d=-\alpha$.  The limit as $d\to 1$ is $\pi |t|$.
\end{proof}

There is an extension to any exceedance measure that is an inverse-power mixture.

\subsection{Identifiability}\label{identifiability}
The convolution-mixture theorem asserts that
\begin{equation}\label{cm}
\CM_d(\alpha, \sigma_0^2) \star N(0, \sigma_1^2) = \CM_d(\rho,\, \sigma_0^2 + \sigma_1^2 ),
\end{equation}
where  $\rho = \alpha\sigma_0^d/(\sigma_0^2 + \sigma_1^2)^{d/2}$.
Of the four parameters $(\sigma_0^2, \sigma_1^2, \alpha, d)$, only three are identifiable
in the marginal distribution.
In principle, lack of identifiability is a serious inferential obstacle because 
the conditional distribution of the signal given $Y=y$ depends on all four parameters.
The difficulty is resolved in this paper, as it is elsewhere in the literature, by fixing $\sigma_1^2 = 1$.
Without such an assumption, there are severe limitations to what can be learned
from the data about the signal.

\section{Conditional distribution}\label{conditionaldistribution}
\subsection{Tweedie's formula}
Provided that the error distribution is standard Gaussian, the argument used in 
\Section~\ref{section:marginaldistribution} shows that the 
moment generating function of the conditional distribution of the signal given $Y=y$ is
\begin{equation}\label{cmgf}
\frac{\phi(y)} {m_\nu(y)}\int_\Real e^{(y+t) u - u^2/2} P_\nu(du) = 
	\frac{1-\rho + \rho \zeta(y+t)} {1-\rho + \rho\zeta(y)},
\end{equation}
depending, to first order in $\rho$, only on the zeta function of the exceedance measure.
In particular, the $r$th conditional moment is $\rho \zeta^{(r)}(y) / (1-\rho + \rho \zeta(y))$,
which is the first-order asymptotic version of Tweedie's formula (Efron, 2011).
For the inverse-square exceedance, the conditional mean is depicted in the middle panel of Figure~1
for a range of sparsity levels.

\subsection{Mixture interpretation}\label{mixtureinterpretation}
Let $G(du; 0)$ be the probability distribution whose moment generating function is $1 + \zeta(t)$,
i.e.,~(\ref{cmgf}) with $y=0$ and $\rho = 1/2$, and
let $G(du; y)$ be the exponentially tilted distribution whose moment generating function is
$(1+\zeta(y + t))/(1+\zeta(y))$, 
i.e.,~(\ref{cmgf}) with $\rho=1/2$.
Then, the  two-component mixture
\begin{equation}\label{G-mixture}
 \frac{1-2\rho}{1-\rho + \rho \zeta(y)} \delta_0(du) + \frac{\rho + \rho\zeta(y)}{1-\rho + \rho \zeta(y)} G(du; y),
\end{equation}
has moment generating function (\ref{cmgf}).
Holding $\rho \zeta(y) = \lambda$ fixed as $\nu \to 0$,
this heuristic argument suggests that the asymptotic conditional distribution of the signal
is a specific two-component mixture in which the $r$th moment of $G$~is
$\zeta^{(r)}(y)/(1+\zeta(y))$.
The argument is non-rigorous because  there is no limit distribution
($\nu \to 0$ for fixed $\lambda > 0$ implies $|y| \to \infty$ and $|\mu| \to \infty$),
but the moment-matching intuition is essentially correct.
However, the Dirac atom in (\ref{G-mixture}) could be replaced by $N(0, \rho^2)$
with no first-order effect on the generating function,
so,  (\ref{cmgf})~does not imply that the conditional distribution has an atom at zero.

For the inverse-power family, the Laplace integral approximation for large $|y|$ is 
\[
\log\bigl(1 + \zeta(y)\bigr) = \vhalf y^2 - (d+1)\log|y|  + \hbox{const} + O(|y|^{-1}),
\]
implying that $G$~is approximately Gaussian with mean $m = y - (d+1)/m$ and unit variance.
Asymptotically,  $\nu \to 0$ for fixed $\lambda > 0$ implies $|y| \asymp \sqrt{2\log(\lambda/\rho)}$,
so the two components in (\ref{G-mixture})
are asymptotically well separated with negligible probability assigned to bounded intervals $(a, b)$
for which $a > 0$.
This behaviour is typical for exceedance measures whose log density is slowly varying at infinity,
but it is not universal.
Bounded and discrete exceedance measures exhibit very different behaviours.

\subsection{Symmetrization}\label{symmetrization}
The conditional density of the signal given $Y=y$ is proportional to the joint density,
\[
\phi(y)\, e^{y u - u^2\!/2} P_\nu(du),
\]
and the symmetrized conditional distribution is proportional to
\begin{eqnarray*}
\cosh(y u) e^{-u^2\!/2}\, P_\nu(du) 
	&=& e^{-u^2\!/2}\, P_\nu(du) + \bigl(\cosh(y u) - 1\bigr) e^{-u^2\!/2} P_\nu(du) \\
	&=& e^{-u^2\!/2}\, P_\nu(du) + \rho \zeta(du; y) + o(\rho).
\end{eqnarray*}
The latter approximation is understood in the usual sense of integrals.

The ratio of the conditional density at $u$ to that at $-u$ is $e^{2 y u}$,
so the conditional density may be recovered from the symmetrized version by
multiplication by the bias function $e^{y u}/\cosh(y u)$.
Without loss of generality, therefore, we focus on the sparse limit of the symmetrized distribution shown above,
ignoring terms of order~$o(\rho)$.

To first order, the symmetrized conditional distribution is a mixture consisting of two components:
\begin{enumerate}
\item{} A central spike distribution $e^{-u^2/2}P_\nu(du)/(1-\rho)$ with weight proportional to $1-\rho$;
\item{} The zeta distribution $\zeta(du; y)/ \zeta(y)$ with weight proportional to $\rho \zeta(y)$.
\end{enumerate}
The moment-generating function of the central spike is $\bigl(1-\rho + \rho \zeta(t)\bigr)/(1-\rho)$,
and the normalization constant for the mixture is the total weight $1-\rho + \rho \zeta(y)$.

For $y^2 \le 3$, the inequality $\zeta(y) \le y^2$ (see \Section~\ref{zetainequalities}) implies that
the net weight on the central spike is at least $1-\rho y^2$. 
In other words, for typical $\rho$-values less than 5\%, and $y^2 \le 3$,
the central spike is the dominant feature of the conditional distribution.

If $\AI(H) > 1$, the zeta density has an integrable singularity at the origin.
Ordinarily, this spike is not visible because it overlaps
the central spike and its relative weight is small.
But there are exceptions in which the central spike has zero density at the origin.
See Fig.~3 in \Section~\ref{illustrations} for one illustration.

\subsection{Signal activity probability}
The conditional distribution given $Y=0$ is symmetric with sparse-limit distribution
\[
P_\nu(du \given Y=0) = (1-\rho)^{-1} e^{-u^2/2} P_\nu(du) + o(\rho),
\]
which is negligibly different from~$P_\nu$.
For any symmetric event such as $A_\epsilon$ whose closure does not include zero,
the first-order conditional probability given $Y = y$ is the weighted linear combination
\begin{eqnarray*} 
P_\nu(|\mu| > \epsilon \given y) 
	&=&
	 \frac{(1-\rho)P_\nu(A_\epsilon \given 0) + \rho \zeta(A_\epsilon; y)} {1-\rho + \rho \zeta(y)} \\
	 &\ge& \frac{\rho \zeta(A_\epsilon; y)} {1-\rho + \rho \zeta(y)}.
\end{eqnarray*}
Since $\zeta(A_\epsilon; y) =\zeta(y) - \zeta(\bar A_\epsilon; y)$, and
(\ref{nozeroatom})~implies that $\zeta(\bar A_\epsilon; y)$ tends to zero for low thresholds,
the asymptotic low-threshold activity bound for fixed~$y$ is
\begin{eqnarray}\label{exceedanceprob}
P_\nu(|\mu| > \epsilon \given y) \ge  \frac{\rho \zeta(y) + o(\rho e^\epsilon) } {1-\rho + \rho \zeta(y)}.
\end{eqnarray}

The argument for (\ref{exceedanceprob}) or its complement as an asymptotic approximation with 
negligible asymptotic error  is more delicate than that for the inequality.
First, the approximation requires $P_\nu(A_\epsilon \given 0) \to 0$ as $\nu \to 0$,
which implies a lower bound $\epsilon \gg\rho$ on the activity threshold.
Second, for $y \to \infty$ at a rate such that $\rho \zeta(y)$ is  bounded below,
the approximation also requires $\zeta(\bar A_\epsilon; y)/\zeta(y)\to 0$.
Ordinarily, if $H$~has unbounded support,
$\log \zeta(y)$ increases super-linearly,  i.e.,~$\lim_{y\to\infty} y^{-1} \log\zeta(y) = \infty$,
in which case the condition that $\rho \zeta(y)$ be bounded below implies 
$e^{\alpha y}/\zeta(y) \to 0$ for every $\alpha$.
Consequently, for every fixed threshold~$\epsilon > 0$, the inequality (\ref{zetaupperbound}) implies
\begin{eqnarray*}
\frac{\zeta(\bar A_\epsilon; y)}{\zeta(y)} 
	&\le& 2 \frac{\cosh(y\epsilon) - 1} {\epsilon^2\, \zeta(y)} .
\end{eqnarray*}
Super-linearity implies that the ratio tends to zero as $\nu \to 0$.
The zero-order conditional non-exceedance probability is then
\begin{eqnarray*}
P_\nu(|\mu| \le \epsilon \given y) &=& 
	\frac{(1-\rho)P_\nu(\bar A_\epsilon \given 0) + \rho\zeta(\bar A_\epsilon; y)}
			{1 - \rho + \rho \zeta(y)} + o(\rho),\\
	&=& \frac{1 - \rho - o(1)} {1-\rho + \rho \zeta(y)},
\end{eqnarray*}
which is independent of~$\epsilon$.
The non-exceedance probability serves as an upper bound for the local false-positive rate, 
\begin{equation}\label{lfpr}
P_\nu(\mu = 0 \given y) \le P_\nu(|\mu| \le \epsilon \given y) = \frac{1- \rho - o(1)}{1-\rho + \rho \zeta(y)}.
\end{equation}

This derivation rests on super-linearity of $\log\zeta(y)$, which 
is satisfied by all inverse-power measures and all of the typical examples discussed in \Section~\ref{Examples}.
Super-linearity fails if $H$~has bounded support,  and, in that case, (\ref{lfpr})~is 
true only for thresholds such that $H(A_\epsilon) > 0$.

\subsection{Tail average activity}
Multiplication of (\ref{lfpr}) by the marginal density, and integration over $y \ge t$ 
gives the tail-average $\epsilon$-inactivity probability
\begin{eqnarray}
\nonumber
m_\nu(y) P_\nu(|\mu| \le \epsilon \given Y=y) &=&  \phi(y) + o(1), \\
\label{BH}
P_\nu(|\mu| \le \epsilon \given Y > t) &=& \frac{1 - \Phi(t) + o(1)} {m_\nu(Y > t) }.
\end{eqnarray}
For bounded $\rho \zeta(t)$ and every fixed threshold $\epsilon > 0$,
this ratio of tail integrals is a re-interpretation of the  Benjamini-Hochberg procedure,
which determines the data threshold $t$ corresponding to any specified tail-average inactivity rate.
The B-H procedure controls the tail-average false-positive rate $P_\nu(\mu=0 \given Y > t)$ 
in the sense that the threshold~$t$ satisfying (\ref{BH}) serves as an upper bound.
It does not approximate the tail-average false-positive rate or the local false-positive rate,
both of which depend crucially on the null atom $P_\nu(\mu = 0)$, which could be zero.

%

\section{Hyperactivity}\label{hyperactivity}
\subsection{The Student~$t$ scale family}
Student's $t_3$ scale family of signal distributions satisfies
\[
\lim_{\nu \to 0} \nu^{-3} P_\nu(dx) = \lim_{\nu \to 0} \nu^{-3} \frac{2\nu^3\, dx} {\pi (\nu^2 + x^2)^2}
	= \frac{2\, dx} {\pi |x|^4} = H(dx),
\]
for $x \neq 0$, so the limit measure exists with rate $\rho_\nu = \nu^3$.
Unfortunately this limit is not a L\'evy measure, so
definition (\ref{exceedancelimit}) is not satisfied, and the  approximations developed in
the preceding sections do not apply.
For example, if $w(x) = \min(x^2, 1)$ or $1 - e^{-x^2}$, the integral $P_\nu(w)$ behaves as $\nu^2 w''(0)$,
i.e.,~$P_\nu(w) = O(\rho^{2/3})$, not $O(\rho)$.

The $t_3$-scale family is an instance of a first-order hyperactive model for which
the exceedance measure exists and $x^2 \min(x^2, 1)$ is $H$-integrable.
First-order hyperactivity typically implies that the exceedance density near the origin is 
$O(|x|^{-d-1})$ for some $2 < d < 4$.
The next section provides a sketch of the modifications needed to accommodate such behaviour.

\subsection{Hyperactivity integrals}
Let $H(dx)$ be a first-order hyperactive exceedance measure, i.e.,~$x^2 H(dx)$ is a non-zero 
symmetric L\'evy measure.
Among the positive multiples of~$H$, the natural reference point satisfies
\begin{equation}\label{h-standardization}
\int_{\Real\setminus\{0\}} \bigl(1 - e^{-x^2/2}(1 + x^2/2) \bigr) \, H(dx) = 1.
\end{equation}
For the $t_3$~model, the unit inverse quartic exceedance density is $3\sqrt{2/\pi}/|x|^4$.

The first-order asymptotic theory for hyperactive sparse models
is determined by the exceedance measure plus two rate parameters
$\gamma_\nu, \rho_\nu$ as follows:
\begin{eqnarray*}
\gamma_\nu &=& \vhalf \int_\Real x^2 e^{-x^2/2} \, P_\nu(dx)  ; \\
\lim_{\nu \to 0} \rho_\nu^{-1} \int_{\Real} x^2 w(x) P_\nu(dx) &=& \int_{\Real\setminus\{0\}} x^2 w(x)\, H(dx)
\end{eqnarray*}
for bounded continuous L\'evy-integrable functions $w \in \W^\sharp$.
The rate parameters for the $t_3$-scale model are $\gamma_\nu = \nu^2\!/2$ 
and $\rho_\nu = \sqrt{2/\pi}\, \nu^3\!/3$, both tending to zero as $\nu \to 0$,
but not at the same rate.

To first order in sparsity,
\begin{eqnarray*}
\int_\Real (1+x^2/2) e^{-x^2/2} P_\nu(dx) 
	&=& 1 - \int_\Real \bigl(1 - e^{-x^2/2}(1 + x^2/2) \bigr) P_\nu(dx) \\
	&=& 1 - \rho \int_{\Real\setminus\{0\}} \bigl(1 - e^{-x^2/2}(1 + x^2/2) \bigr) \, H(dx) \\
	&=&  1-\rho,
\end{eqnarray*}
implying that $\int_\Real e^{-x^2/2}\, P_\nu(dx) = 1-\rho - \gamma$.

To ensure integrability at the origin, the definition of the zeta function is modified to
\[
\zeta(t) = \int_{\Real\setminus\{0\}} \bigl(\cosh(tx) - 1 - t^2 x^2/2\bigr) e^{-x^2/2}\, H(dx),
\]
implying that $\zeta(t) = O(t^4)$ near the origin.
Provided that $H$ satisfies (\ref{h-standardization}), the product $\psi(x) = \phi(x) \zeta(x)$ is a probability density function.
Its characteristic function is
\[
\int_\Real e^{itx} \psi(x)\, dx = e^{-t^2/2}  + e^{-t^2/2} \int_{\Real\setminus\{0\}} \bigl(\cos(tx) -1 + t^2 x^2/2 \bigr)\, H(dx),
\]
which simplifies to
$e^{-t^2/2}(1 +  |t|^3 \sqrt{\pi/2})$
for the inverse quartic.

The marginal distribution of $Y$ for a first-order hyperactive sparse model is a three-component mixture
of the density functions $\phi(y)$, $y^2 \phi(y)$ and $\psi(y)$:
\[
m_\nu(y) = (1-\gamma - \rho) \phi(y)   + \gamma y^2\phi(y) + \rho \psi(y)+ o(\rho).
\]
Note that the basis distributions are fixed, while the coefficients $\gamma_\nu, \rho_\nu$ are sparsity-dependent.
The chief consequence of signal hyperactivity is that the non-Gaussian perturbations 
are not of equal order in sparsity: asymptotically, $\gamma_\nu > \rho_\nu$.
If it is convenient, the first two components may be combined so that
\[
m_\nu = (1 - \rho)N(0, 1+2\gamma) + \rho \psi + o(\rho),
\]
at the cost of a small increase in the variance of the Gaussian component.

All of the results in \Sections~3 and~4 may be extended to hyperactive models
with $\rho \zeta(y)$ replaced by $\gamma y^2 + \rho \zeta(y)$.
Tweedie's formula for the conditional mean of the signal involves both rate parameters:
\[
E(\mu \given Y=y) = \frac{2\gamma y + \rho \zeta'(y)} {1-\gamma-\rho + \gamma y^2 + \rho \zeta(y)}.
\]

\section{Illustration}\label{illustrations}
The left panel of Figure~2 shows a histogram of the absolute values of 5000 independent responses
generated by Efron's (2011) version of the sparse signal plus Gaussian noise model
\[
Y_i = \mu_i + \varepsilon_i
\]
where the $\varepsilon$s are independent $N(0, 1)$ random variables,
and the signals are
\[
\mu_i = \pm \log((i - 1/2)/500)
\]
for $i \le 500$, and $\mu_i = 0$ otherwise.
The absolute non-zero signals are approximately exponentially distributed,
and the mixture fraction is 10\%.
But a substantial fraction of the signals are small, and consequently  undetectable in the presence of
additive standard Gaussian noise.
Example~2 in \Section~\ref{Examples} implies that the effective mixture fraction is 
\[
\rho = 0.1 \times \vhalf \int (1 - e^{-x^2/2}) e^{ - |x|} \, dx \simeq  0.0344.
\]

\begin{figure}
\centerline{
\includegraphics[scale = 0.40]{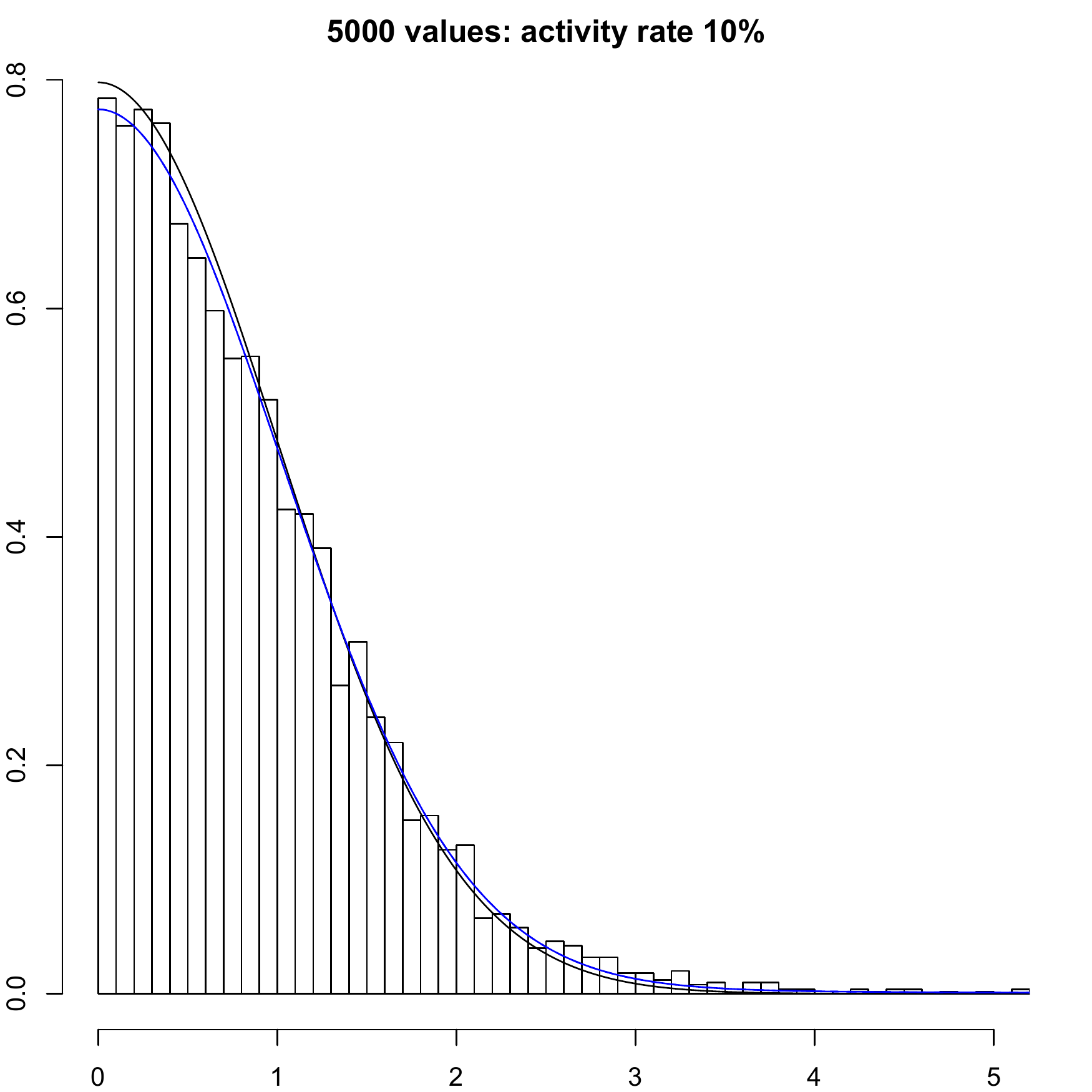}\hfil
\includegraphics[scale = 0.40]{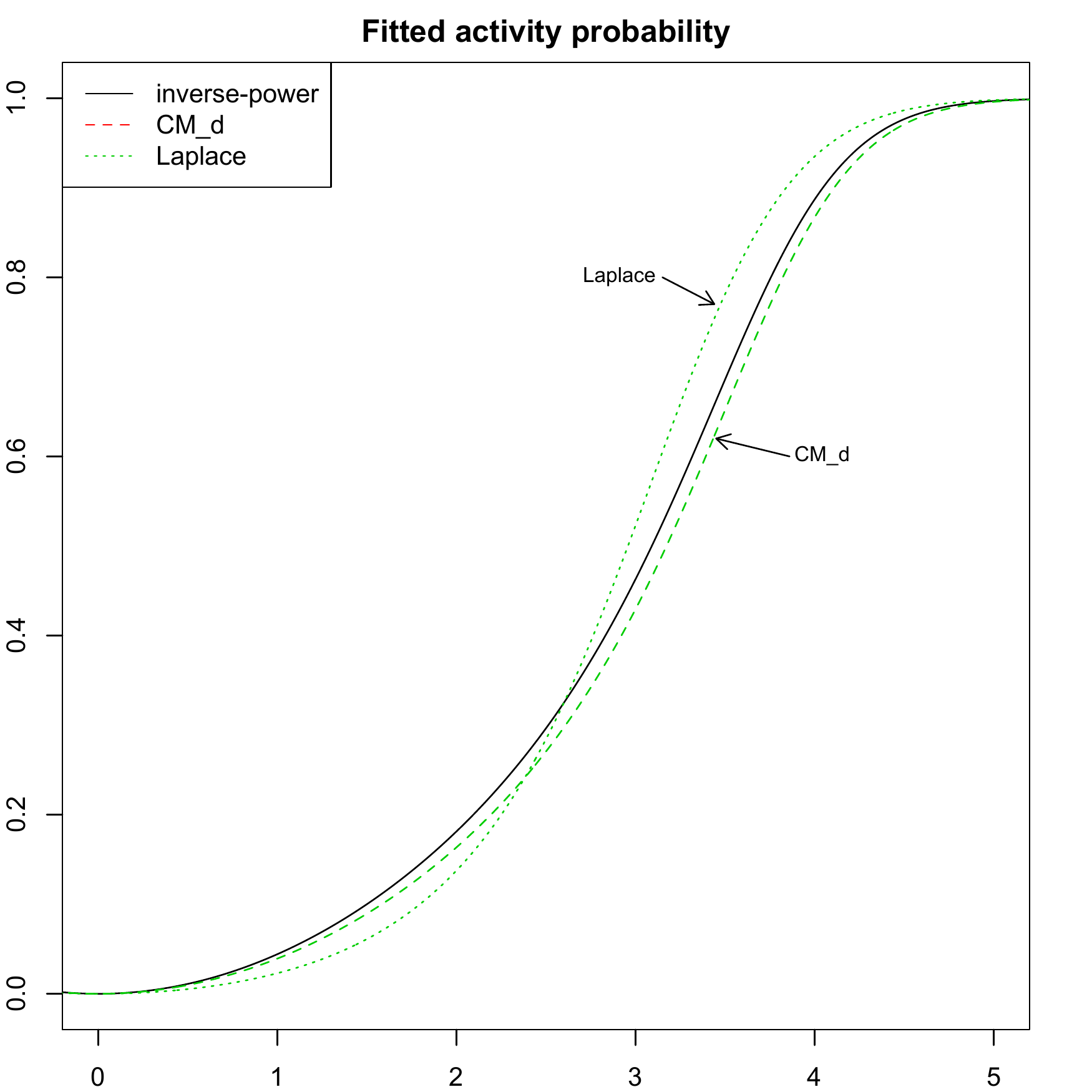}
}
\caption{Fitted marginal density and fitted signal activity probability as a function of~$y$.}
\end{figure}

The developments  in this paper suggest two ways to proceed,
both using the inverse-power family of exceedance measures for illustration.
The first is to estimate the parameter $(\rho, d)$ by maximizing the asymptotic log likelihood
\[
l(\rho, d; y) = \sum \log\bigl(1 - \rho + \rho \zeta_d(y_i) \bigr).
\]
Maximization with no constraints on~$\rho$ gives $\hat d = 1.49$, $\hat\rho = 0.056$, 
and $l(\hat\rho, \hat d) = 123.32$ relative to the value at $\rho = 0$.
The solid line in Fig~2b shows the fitted conditional activity probability
$\rho \zeta_d(y)/(1-\rho+\rho \zeta_d(y))$ as a function of~$y$.

The preferred option is to include a free scale parameter, $\sigma^2 = 1+\sigma_0^2$, and to estimate 
subject to the condition $\rho \le (\sigma_0/\sigma)^d$ as implied by the convolution-mixture~(\ref{cm}).
The log likelihood function for the $\CM_d(\rho,\sigma^2)$ model is
\[
- \sum y_i^2/(2\sigma^2) - n\log\sigma +
 \sum \log\bigl(1 - \rho + \rho \zeta_d(y_i/\sigma) \bigr) .
 \]
Constrained maximization gives
\[
\hat d = 1.48,\quad \hat\rho = 0.051,\quad \hat\sigma_0 = 0.135,
\]
on the boundary at $\rho=(\sigma_0/\sigma)^d$, for a maximum of 122.95 relative to $\rho = 0$.

For the marginal distribution of absolute values, 
the table below compares five quantiles of the Laplace-Gaussian mixture with
the corresponding quantiles of the fitted $\CM_d$ distribution:
\[
\arraycolsep=5pt
\begin{array}{cccccc}
	 &97\% & 98\%      &99\%    &99.5\%   &99.75\% \\
\hbox{L-G} &2.39 &2.62 & 3.04 & 3.56 & 4.20 \\ 
\CM_d &2.40 & 2.61 & 3.01 & 3.61 & 5.00 
\end{array}
\]
The match is reasonably satisfactory at least up to the 99.5 percentile.
Only at the most extreme quantiles does the difference between the exponential tail
of the L-G mixture and the inverse-power tail of the $\CM_d$ mixture become apparent.
This discrepancy could be viewed as a deficiency of the class of inverse-power measures,
but we are more inclined to view it as a deficiency of the Laplacian model for signals.

%

The dashed line in Fig~2b shows the fitted conditional activity probability
as a function of~$y$.
The difference between the two activity curves is small, and is due partly to the re-scaling 
($\hat\sigma=1.01$) that occurs in the $\CM_d$ fit, and partly to the small difference in fitted rates.
For example, the fitted conditional activity probabilities given $Y=3$ are 46.3\% and 42.9\% respectively.

For this example, we know that the signals were effectively generated using the short-tailed
atom-and-slab Laplace model, which is associated with the two-parameter L\'evy measure 
$H_{\lambda,\tau}$ in~(\ref{laplacelevy}) with $\tau=1$ and $e^{-\lambda}=0.9$.
Using the associated zeta function, the asymptotic log likelihood achieves a maximum of 135.3 at $\hat\tau=1.00$, $\hat\rho = 0.043$.
In this setting, $\rho$~is the L\'evy convolution parameter, 
and the likelihood function is essentially constant in $\lambda$ over the range $0 \le \lambda \le 0.5$.
For $\lambda \to 0$, the marginal density (\ref{marginalmixture}) covers both atomic and non-atomic
spike-and-slab Laplace models such as $(1-\nu)e^{-|x|/\nu}/(2\nu) + \nu \tau e^{-\tau |x|}/2$, and
the log likelihood is the same for all models in this equivalence class.
The Laplace-activity curve $\rho \zeta(y)/(1-\rho + \rho \zeta(y))$ shown in Fig.~2b also applies in the non-atomic setting,
with the threshold-exceedance interpretation (\ref{lfpr}).


\begin{figure}
\centerline{
\includegraphics[scale = 0.5]{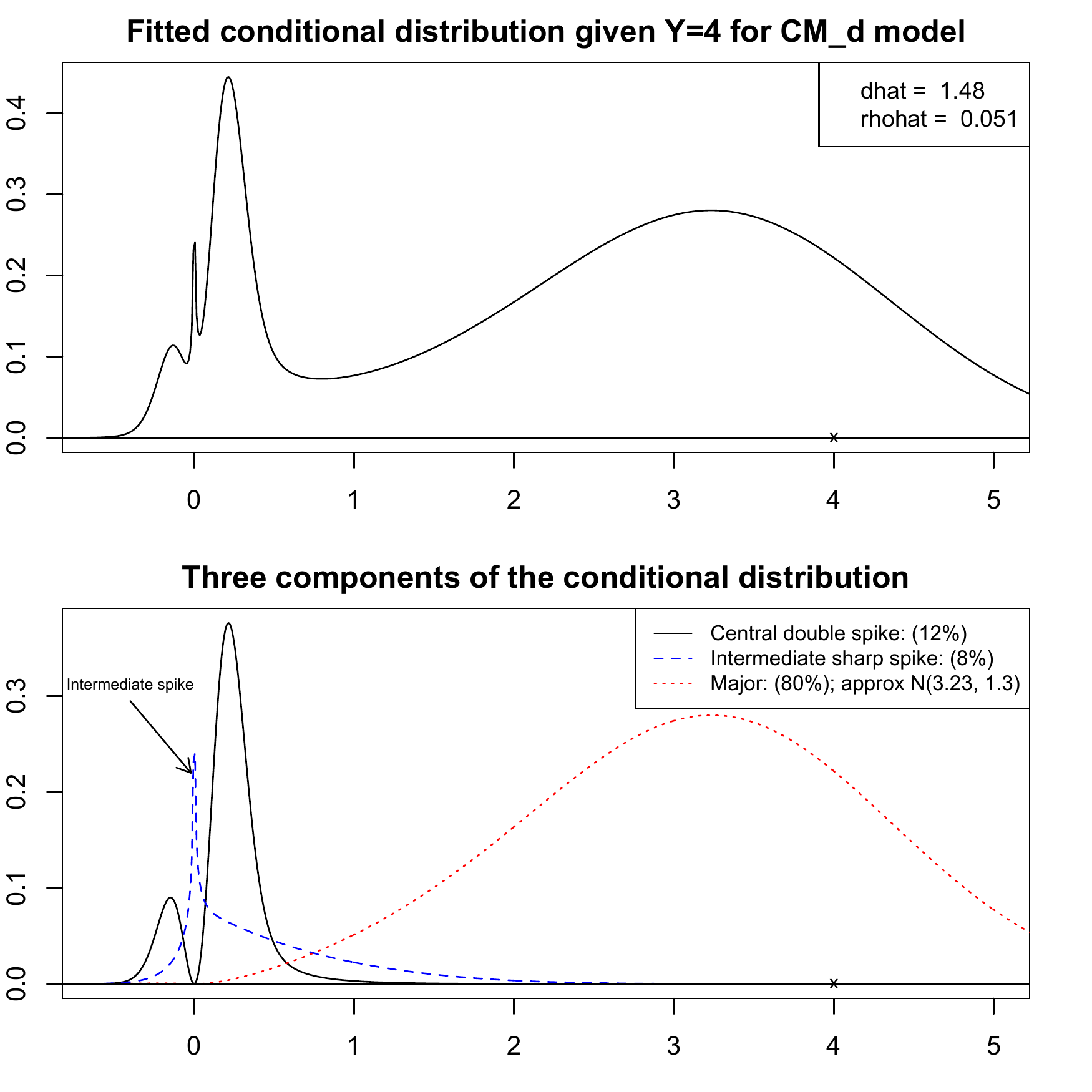}\hfil
}
\caption{Fitted conditional distribution showing the singularity at the origin in the $\CM_d$ model.
The lower panel shows the decomposition as a three-part mixture.}
\end{figure}

In the $\CM_d$ model, $\AI(\hat H) = \hat d > 1$ implies that the fitted
conditional density given $Y=y$ has a $|u|^{1-d}$-singularity
at the origin, which is clearly visible for $y=4$ in Fig.~3a.
Figure~3b shows the additive decomposition of the conditional density in which 
the central double spike has net weight~12\%, and the zeta component has weight 88\%.
The zeta-measure is decomposed further as a two-part mixture
along the lines of~\Section~\ref{hyperactivity}.
The intermediate spike has density $u^2 e^{-u^2/2} H(du) / \zeta_2$ and weight
proportional to $\rho \zeta_2 y^2/2$, which is asymptotically negligible compared with $\rho \zeta(y)$, 
but not numerically negligible for $\rho=0.051$.
The remaining major component is unimodal with density
\[
\frac{\zeta(du; y)  -  y^2 u^2 e^{-u^2/2} H(du)/2} {\zeta(y) -\zeta_2 y^2/2},
\]
and weight proportional to $\rho\zeta(y) - \rho \zeta_2 y^2/2$.
The latter can be approximated with reasonable accuracy by a Gaussian distribution.

The asymptotic theory in this paper tells us that, if $\rho$ is small, 
every model in the inverse-power class, such as the sparse Student~$t$,
must produce essentially the same fit, with similar estimates for $(\rho, d)$.
In all cases, the conditional distribution given $Y=y$ has a two-part decomposition as
described in \Section~\ref{symmetrization}.
The asymptotic theory says little about the appearance of the central spike
other than its total mass and the fact that it is concentrated near the origin.
The central spike is bimodal for the $\CM_d$ model but
unimodal for the sparse Student~$t$ and most atom-free spike-and-slab mixtures.
In neither case could the conditional distribution be said to have an atom at zero.
But the asymptotic theory also tells us that the zeta component depends only on the exceedance measure,
so the zeta measure,
and its two-part decomposition in Figure~3, are the same for all models in the same sparsity class.
To that extent at least, only the exceedance measure matters.
There are indications that a second-order analysis might be capable of 
offering a more detailed description
of the behaviour of the conditional distribution in the neighbourhood of the origin,
but this paper stops at first order.

%

\section{Acknowledgements}
We are grateful to Jiyi Liu for pointing out that the convolution-mixture property in section~\ref{section:convolutionmixture}
implies a certain form for the characteristic function, and ultimately for supplying a proof
that the characteristic function of $\psi$ has this form.

\section{References}
\everypar={\parskip 0.5pc \parindent 0pt \hangindent=12pt}
\noindent 
%
Barndorff-Nielsen, O.E. and Hubalek, F. (2008)
Probability measures, L\'evy measures and analyticity in time.
\textit{Bernoulli} 14, 764--790.

Benjamini, Y., and Hochberg, Y.  (1995). 
Controlling the false discovery rate: A practical and powerful approach to multiple testing. 
\textit{J.~Roy.\ Statist.\ Soc.} B, 57, 289-300


Carvalho, C.M., Polson, N.G. and Scott, J.G. (2010)
The horseshoe estimator for sparse signals.
{\it Biometrika} 97, 465--480.

Davison, A.C. and Smith, R.L. (1990)
Models for exceedances over high thresholds.
\textit{J.~Roy.\ Statist.\ Soc.}~B 52, 393--442.

Efron, B. (2008)
Microarrays, empirical Bayes and the two-groups model.
\textit{Statist. Sci.} 23, 1--22.

Efron, B. (2009)
Empirical Bayes estimates for large-scale prediction problems.
\textit{Journal of the American Statistical Association}, 104, 1015--1028.

Efron, B. (2010) 
{\it Large-Scale Inference:  Empirical Bayes Methods for Estimation, Testing and Prediction.}
Cambridge University Press.

Efron, B. (2011)
Tweedie's formula and selection bias.
\textit{Journal of the American Statistical Association}, 106, 1602--1614.


George, E.I. and McCulloch, R.E. (1993)
Variable selection via Gibbs sampling.
\textit{J.~Amer.\ Statist.\ Assoc.}, 88, 881--889.

Griffin, J.E. and Brown, P.E. (2013)
Some priors for sparse regression modelling.
\textit{Bayesian Analysis} 8, 691--702.

 
Johnstone, I. and Silverman, B.W. (2004)
Needles and straw in haystacks: Empirical-Bayes estimates of possibly sparse sequences. 
\textit{The Annals of Statistics}, 32, 1594--1649.

Johnson, V.E. and Rossell, D. (2010) 
On the use of non-local prior densities in Bayesian hypothesis tests.
{\it J.~Roy. Statist.\ Soc.}~B  72, 143--170.


Rockova, V. and George, E.I. (2018)
The spike and slab LASSO.
\textit{Journal of the American Statistical Association} (to appear).
https://doi.org/10.1080/01621459.2016.1260469

\end{document}